\documentclass{article}

\usepackage{arxiv}

\usepackage[utf8]{inputenc} % allow utf-8 input
\usepackage[T1]{fontenc}    % use 8-bit T1 fonts
\usepackage{url}            % simple URL typesetting
\usepackage{booktabs}       % professional-quality tables
\usepackage{amsfonts}       % blackboard math symbols
\usepackage{nicefrac}       % compact symbols for 1/2, etc.
\usepackage{microtype}      % microtypography
\usepackage{lipsum}		% Can be removed after putting your text content
\usepackage{graphicx}
\usepackage{natbib}
\usepackage{doi}
\usepackage{xr}
\usepackage{xcite}
\usepackage{soul}
\usepackage{amsmath}
\usepackage{gensymb}
\usepackage{amssymb}
\usepackage{marvosym}

\title{Enhanced Impact Mitigation via 3D-Multilayered Material Architectures}

\author{{Thomas Butruille} \\
	Department of Mechanical Engineering\\
	Massachusetts Institute of Technology\\
	Cambridge, MA 02139 \\
	\texttt{thomasb3@mit.edu} \\
	\And
	{Joshua C.~Crone} \\
	Physical Modeling and Simulation Branch\\
	DEVCOM Army Research Laboratory\\
	Aberdeen Proving Ground, MD 21005 \\
	\texttt{joshua.crone.civ@army.mil} \\
    \And
	{Carlos M.~Portela$^*$} \\
	Department of Mechanical Engineering\\
    Institute of Soldier Nanotechnologies\\
	Massachusetts Institute of Technology\\
	Cambridge, MA 02139 \\
	\texttt{cportela@mit.edu} \\
}

\hypersetup{
pdftitle={Enhanced Impact Mitigation via 3D-Multilayered Material Architectures},
pdfauthor={Thomas Butruille, Joshua C.~Crone, Carlos M.~Portela},
pdfkeywords={architected materials \and metamaterials \and impact \and energy dissipation \and LIPIT},
}

\begin{document}
\maketitle

\begin{abstract}
	 Materials designed by nature commonly exhibit functional grading and laminated structures, particularly when intended for enhanced impact protection. Synthetic materials have also found success in exploiting this concept with fully dense but spatially varying architectures, as is the case with advanced fiber-based composites. In the lightweight materials space, porous architected materials have shown benefits for extreme impact mitigation, proving to be advantageous in dissipating large amounts of energy per unit mass, but rarely harness the benefits of layering or functional grading in designs. Here, a design paradigm for lightweight multilayered materials towards high impact-mitigation efficacy is demonstrated, showing that the use of alternating monolithic and beam-based architectures leads to enhanced and predictable responses under extreme conditions. These layered, mass-equivalent `heterostructures' with different ordering and proportions of octet and monolithic layers outperform single-architecture lattices on a mass-normalized energy dissipation basis by $>$50\% when subjected to supersonic microparticle impact. Through analysis that combines wave-propagation analysis, nonlinear finite element simulations, and post-impact crater reconstruction, layer-by-layer mechanical properties are mapped to crater formation and energy dissipation behaviors. This heterostructure design framework offers a simple approach towards tuning failure and impact resistance of materials for protective applications from Whipple shields to sports equipment. 
\end{abstract}


\section{Introduction}
Impact mitigation has been a topic of research interest for centuries, with conventional strategies leveraging available materials to distribute stress and dissipate energy. This interest has been a natural extension of the critical evolutionary pressure of protection through the use of biological materials. A multitude of biomaterial systems have evolved to offer protection from repeated or extreme impact, such as the porous, hierarchical bone structure through the beak and skull of a woodpecker \cite{Lee2014Woodpecker}, the multi-material brick-and-mortar structure of nacre \cite{Luz2009, Islam2021}, and the Bouligand structure in the dactyl club of the mantis shrimp \cite{Suksangpanya2017,Behera2021, Alderete2025}. Each of these materials offers a different approach towards impact mitigation, but have withstanding commonalities in the presence of hierarchy and gradation in density and structure. Humans have taken inspiration from these systems in the layering of different protective materials such as metal plates, chain mail, cloth, leather \cite{ElMessiry2021, Skaggs2003, Ramdayal2016} and even the packing of sand \cite{Song2009, Katsuragi2007, Omidvar2014} for ballistic impact. 

Historical to contemporary protective systems---ranging from metallic armor to laminated composites---tend to be fully dense and still largely achieve strength and impact resistance by homogeneously adding mass rather than by distributing it strategically. This constraint has motivated the development of lightweight alternatives such as aerogels \cite{Tan2001} and sandwich structures \cite{Mines2013, Evans2010, Yungwirth2008}, culminating in designs across loading rates like the Whipple shield for micrometeoroid protection \cite{Wen2021}.  More recently, 3D architected materials, i.e., materials with engineered 3D nano/microstructures, have expanded the structure-property map of engineered materials \cite{Ashby2006, Bauer2017, Zhang2020, Surjadi2025}, offering unprecedented control over mechanical performance at multiple length scales. Combining advanced materials such as polymers, ceramics, and alloys with intentionally designed architecture has enabled extreme stiffness- and strength-to-weight ratios \cite{Singh2024, MezaGreer2014, Jang2013, Schaedler2011}, with future potential for tunable resistance to fracture \cite{O'Masta2017, Shaikeea2022, Fulco2025, Karapiperis2025}. Drop-tower and pressure bar experiments to loading strain rates of 10$^3$ s\textsuperscript{-1} in stochastic \cite{Deshpande2000, Barnes2014, Gaitanaros2014, Gaitanaros2015} and periodic shell- \cite{Novak2023, Tancogne-Dejean2019} and beam-based architectures \cite{Mines2013, WeeksOctet2022, Weeks2023} have provided experimental evidence of the influence of architecture on dynamic deformation mechanisms for enhanced energy dissipation such as shock formation. Studies on the drop-weight and ballistic impact of composites \cite{Naik2004}, laminates \cite{Lee2001, Shyr2003}, or sandwich structures \cite{Yungwirth2008, Zhou2013, Ma2021} at similar strain rates have examined how complex dynamic loading conditions and the interplay of failure in a multi-material system engender inertial and fracture-based energy dissipation components. Although the properties of architected materials subject to quasistatic loading are widely known, the design space for these materials in dynamic and extreme loading conditions remains underexplored due to fabrication and experimental challenges in this regime, as combined fabrication and characterization times for single drop-tower or pressure bar experiments can stretch from several hours to days. 

Direct-impact setups \cite{Hawreliak2016, Dattelbaum2020} on microscale samples improve experimental throughput by reducing fabrication times while simultaneously increasing strain rates towards 10$^{5}$ s\textsuperscript{-1}. Laser-induced particle impact testing (LIPIT) pushes strain rates towards $\sim$10$^7$ s\textsuperscript{-1} \cite{Lee2014, Hyon2018, Veysett2021} while investigating additional drivers of energy dissipation during impact loading beyond those present in simple uniaxial compaction. Under particle-impact conditions of multilayer graphene \cite{Lee2014}, the strain energy from the impact region is delocalized through the high in-plane wave speed of graphene, which results in petalling deformation that dissipated energy from impacting supersonic particles. Recently, these types of experiments have demonstrated benefits of 3D architecture in activating different mechanisms and mitigating particle impacts in carbon and polymer truss lattices, surpassing the performance of equal-mass monolithic materials~\cite{Portela2021}. Specifically, it has been demonstrated that octet and tetrakaidecahedron truss-lattices are more efficient at dissipating impact energy on a per-unit-mass basis \cite{Butruille2024} compared to a non-architected monolithic layer, despite the monolithic layer increasing the amount of material that participates in the impact event. These investigations offer individual approaches towards impact mitigation in contrast to how biological protective systems employ hierarchy and gradation for multi-modal energy dissipation. For example, the mantis shrimp's dactyl club combines microparticle agglomeration, labyrinthine cracking, and shear-wave filtering at varying depths through its structure to withstand the repeated shocks it generates \cite{Suksangpanya2017, Behera2021, Alderete2025}. Leveraging functionally graded or multi-architecture material systems appears to be a promising avenue for materials design towards increased efficiency in multi-modal energy dissipation.

%%% Figure 1 %%%%%%%%%%%%%%%%%%%%%%%%%%%%
\begin{figure*}[t]
    \centering
    \includegraphics[width= 16 cm]{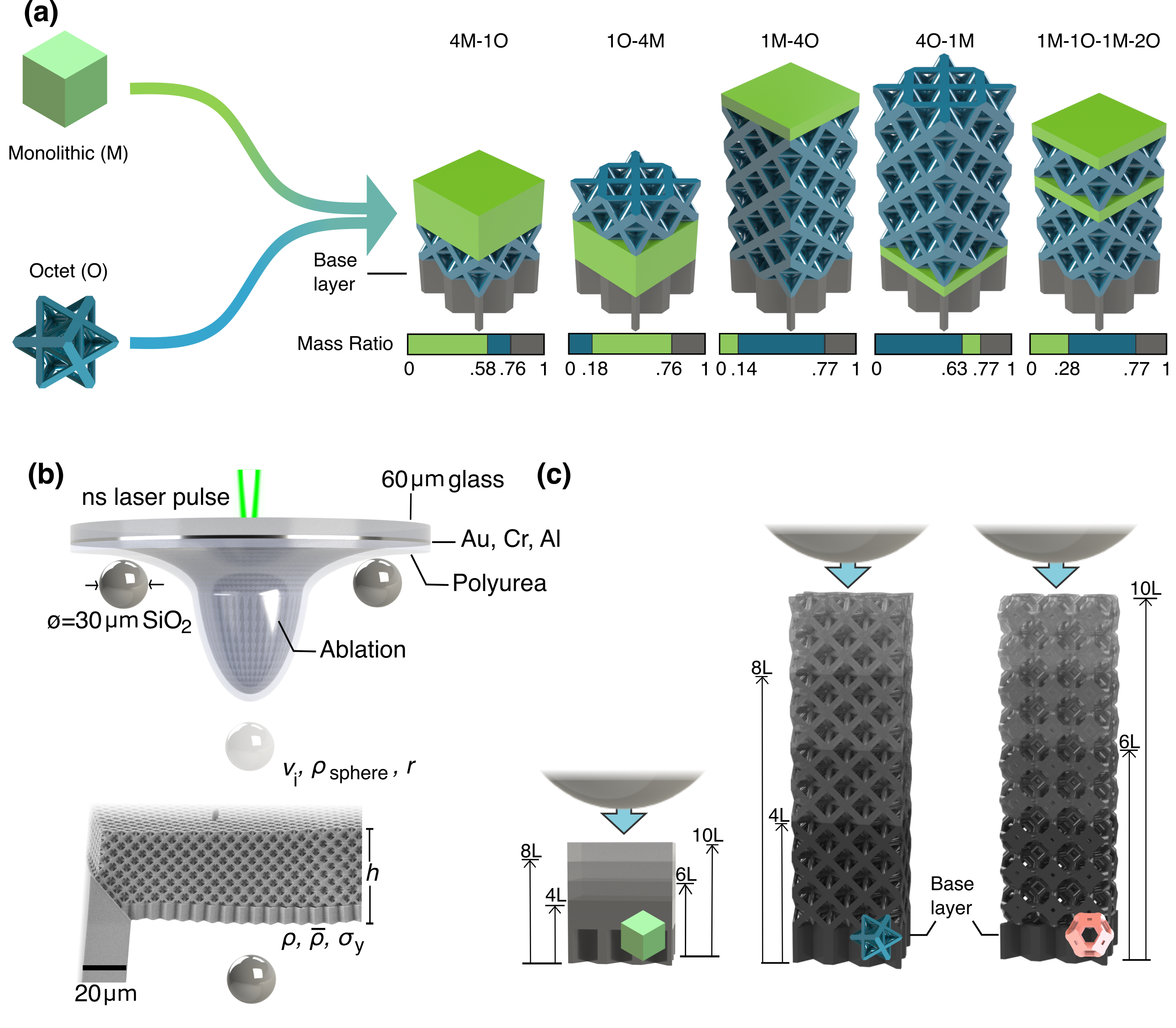}
    \caption{ \textbf{a)} Design framework for heterostructured suspended shields, depicting 4Monolithic (M)-1Octet (O), 1O-4M, 1M-4O, 4O-1M, and 1M-1O-1M-2O architectures and their respective mass ratios by architecture, including a base layer. \textbf{b)} Laser-induced microparticle impact of a suspended lattice and \textbf{c)} monolithic, octet, and tetrakaidecahedron lattices discretized in mass-equivalent layers ($n$L), dictated by shield thickness or tesselation, with $n$ ranging from 4 to 10.}
    \label{fig:DesignParadigm}
\end{figure*}
%%%%%%%%%%%%%%%%%%%%%%%%%%%%%%%%%%%%%%%%%%

In this work, we propose and demonstrate a type of heterostructured architected shield---i.e., materials with multilayered and dissimilar architectures through thickness---which leverage functional grading of architected and non-architected components to increase impact-mitigation efficiency via programmable material recruitment during impact. We perform supersonic microparticle impact experiments, through which we assess differences in impact energy dissipation across architectures, and employ x-ray computed tomography to provide insight into architecture-dependent cratering of heterostructured shields. Using explicit finite element simulations of impact in our suspended architectures, we further identify differences in lateral material recruitment during particle impact, evidenced by analysis of stress-wave propagation versus particle penetration time. Using these experimental and computational insights, we propose an analytical tool of wave propagation to predict crater formation patterns in heterostructured layered materials. Together, our findings establish key design principles for 3D-multilayered material systems for impact conditions, offering new strategies for impact mitigation that mimic principles found in natural multilayered materials.

%/////////////////////////////////////////////////
\section{Microparticle Impact of Multilayered Suspended Lattices}
Taking inspiration from spatially graded biological impact-mitigating systems \cite{Luz2009, Islam2021, Suksangpanya2017, Behera2021, Alderete2025}, which are characterized by features and compositions that vary in three dimensions, a promising direction for particle-impact mitigating materials is the use of heterogeneous microstructures. Heterostructured lattices, which incorporate mixed architected and monolithic layers (\textbf{Figure~\ref{fig:DesignParadigm}}a), separately incorporate various deformation mechanisms ranging from efficient compaction and localized shear-induced fracture in beam-based architectures, to  delocalized radial cracking in monolithic shields \cite{Lee2014, Butruille2024}. To identify optimal distributions of architecture and monolithic material in the design of heterostructured architected materials, we first aimed to determine the effect of thickness on the impact response of each lattice type as well as the monolithic constituent polymer (Figure~\ref{fig:DesignParadigm}b,c). As demonstrated in impact explorations of thin plates \cite{Goldsmith1971, Goldsmith1978}, varying the thickness of these samples (or in our case the number of architected layers) leads to changes in deformation response. An understanding of how impact performance in different architectures scales with thickness (or areal density) is thus essential in designing functionally graded heterostructures.
%%% Figure 2 %%%%%%%%%%%%%%%%%%%%%%%%%%%%
\begin{figure*}[!t]
    \centering
    \includegraphics[width= 16 cm]{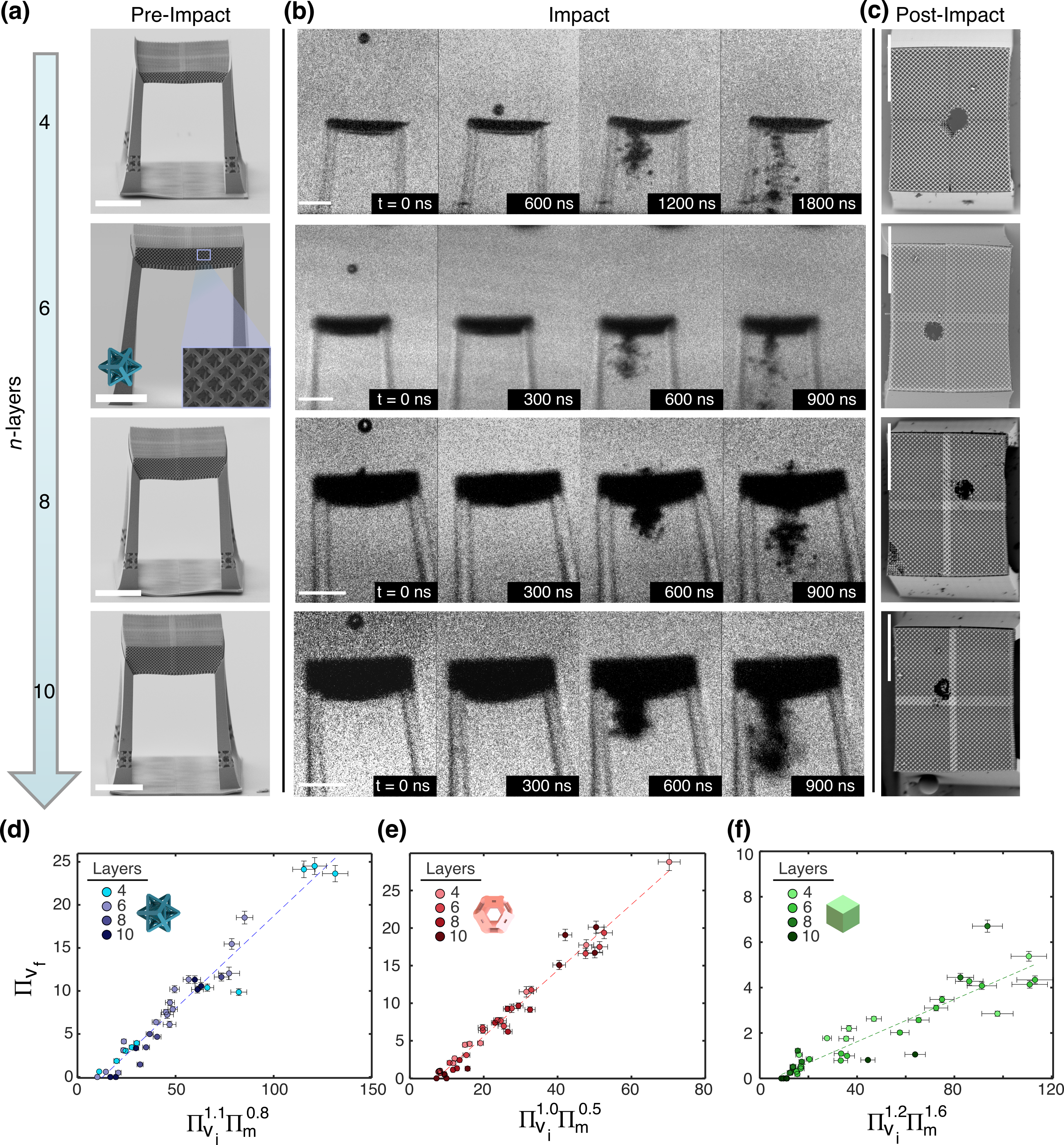}
    \caption{\textbf{a)} Multilayered octet suspended lattices prior to impact, and \textbf{b)} high-speed imaging frames from microparticle impact events at four, six, eight, and ten unit cell thickness (including the base layer), with impact velocities of 367, 562, 792, and 775 m s\textsuperscript{-1} respectively.  \textbf{c)} Post-impact SEM of octet lattice craters across shield thicknesses. \textbf{d-f)} Best-fit dimensional analysis relation between select dimensionless groups for octet, tetrakaidecahedron, and monolithic suspended shields across shield thicknesses. Scale bars, 100 \textmu{}m.}
    \label{fig:DimAnalysis}
\end{figure*}
%%%%%%%%%%%%%%%%%%%%%%%%%%%%%%%%%%%%%%%%%%

To this end, we fabricated a series of suspended shields of varying thickness and architecture through the use of two-photon lithography (Nanoscribe PPGT2) out of IP-Dip photoresist. Each suspended architecture was fabricated in a layer-by-layer fashion onto two adjacent support structures spaced 195 \textmu{}m apart. The first layer consisted of a support base layer (Experiments and Methods) approximately 260 \textmu{}m above silicon substrates, while subsequent architected layers of tetrakaidecahedron or octet lattice architectures were fabricated in tessellations (including the first layer) of 4, 6, 8, and 10 layers (\textbf{Figure~\ref{fig:DimAnalysis}}a, Figure S1). Each architected layer corresponded to a 32$\times$26$\times$1 tessellation of 6.7$\pm$0.1 \textmu{}m cubic unit cells, with relative density (i.e., fill fraction) of $\bar \rho=$33$\pm$1\%. To compare against an equal-mass sample without 3D architecture, we fabricated monolithic suspended plates with thickness (excluding the base layer) ranging from 4.6$\pm$0.7 to 14$\pm$2 \textmu{}m (Figure~\ref{fig:DesignParadigm}c, Figure S2). All samples were fabricated with identical printing parameters, to ensure the same level of crosslinking in the constituent polymer. 

Prior work on microparticle impact of beam-based lattices identified superior energy dissipation metrics for octet architectures compared to tetrakaidecahedron architectures, owing to a likely higher dissipation contribution from creating a cylindrical fracture boundary at the particle impact site \cite{Butruille2024}. However, these experiments were not able to identify how variations of the impact conditions, such as changing the ratio between the impacting particle radius $r$ and the shield thickness $h$, affected the energy dissipation metrics.  
To address this question, we performed microparticle impact experiments on lattices and monolithic shields using the laser-induced particle impact test (LIPIT) method \cite{Hyon2018, Veysett2021}, with $\sim$30 \textmu{}m silica microspheres as projectiles (Figure~\ref{fig:DesignParadigm}b). The incident velocity of these particles ranged from 250 m s\textsuperscript{-1} to $\sim$900 m s\textsuperscript{-1} (Figure S3), with Figure~\ref{fig:DimAnalysis}b and~\ref{fig:DimAnalysis}c depicting the impact event across suspended octet lattices of varying thickness and SEM images of the post-impact craters. Systematically varying shield thickness across architectures thus allowed refinement of a dimensional analysis framework based on the ratio between the particle radius and the shield thickness $r/h$. 

\subsection{Dimensional Analysis}
Dimensional analysis has been used to successfully compare self-similar impact events across several orders of magnitude of impactor size and strain rates \cite{Holsapple1987, Holsapple1993}, from asteroid collisions to water droplets \cite{Zhao2015}, and more recently to LIPIT of microlattices \cite{Portela2021, Butruille2024}.  Traditionally, dimensional analysis of this kind was employed to predict the cratering efficiency of an impact event characterized by dimensionless group $\Pi_V = \rho V/m$, where $\rho$ is the impacted material density, $V$ the crater volume, and $m$ the impactor mass. Here, the impactor was assumed to have zero final velocity ($v_f=0$), with all initial kinetic energy associated with the incident velocity $v_i$ being consumed during crater formation. However, when lattices of finite thickness $h$ are impacted, the nonzero $v_f$ necessitates new dimensionless groups. Using the Buckingham-Pi theorem, we employ a generalized scaling law relating $v_f$ to all relevant impact parameters $v_f=f(\{r, \rho_{\mathrm{sphere}}, v_i\}, \{\sigma_y, \rho, \bar\rho, h\})$, where $r$  and $\rho_{\mathrm{sphere}}$ are the particle radius and density, $\sigma_y$ and $\rho$ are the impacted material's yield strength and density, respectively, and $\bar \rho$ is its relative density. Applying the Buckingham-Pi theorem on these variables leads to five dimensionless groups

\begin{equation}
\label{eq:finalDA}
\begin{split}
    \underbrace{\frac{\rho v_f^{2}}{\sigma_y}=f(\frac{\sigma_y}{\rho v_i^2}, \frac{r}{h}\frac{\rho_{\mathrm{sphere}}}{\rho\bar{\rho}}, \frac{\rho_{\mathrm{sphere}}}{\rho},\bar\rho)} \\
    \Pi_{v_f}=f(\Pi_{v_i},\Pi_m,\Pi_\rho,\bar\rho ).
\end{split}~.
\end{equation}

Here, $\Pi_{v_i}$ and $\Pi_{v_f}$ correspond to the ratios between the sample yield strength and the incident and final dynamic pressures, respectively, while $\Pi_m$ represents the mass ratio between the impactor and the through-thickness localized interaction mass with radius $r$. Approximating the particle impact event as a point source, these groups are arranged into a power law form $\Pi_{v_f} \propto \Pi_{v_i}^\alpha\Pi_m^\beta$ where the exponents $\alpha$ and $\beta$ capture the scaling effects of impact velocity and mass ratio across architectures. Other groups are excluded as they do not change across experiments for a given suspended shield architecture.

Prior work using this dimensionless relation sought to provide a scale-insensitive prediction for impact behavior in architected materials \cite{Butruille2024} but two key limitations emerged. First, all impact data was fit together across octet, tetrakaidecahedron, and monolithic architectures. However, energy dissipation during impact is dependent on two competing terms---the fracture boundary that scales with impactor radius $r$ and the uniaxially compacted region that scales with impactor cross-sectional area or $r^2$. Since the dimensionless group $\Pi_m \propto r/h$ varies linearly with $r$, it should scale differently across material systems where the energy dissipation terms change in relative magnitude. Second, the $\Pi_m$ group being fit only to small variations of particle radius in past analyses made it sensitive to experimental noise. 

To address these limitations, we systematically fit the power law form of Equation~\ref{eq:finalDA} independently for each suspended shield architecture type (i.e., octet, tetrakaidecahedron, and monolithic), solving for $\alpha$ and $\beta$ by individually varying each within the parametric limits of [0,2]. The imposed differences in shield thickness strongly fit the $\Pi_m$ dimensionless parameter, with $\alpha$=[1.1, 1.0, 1.2] and $\beta$=[0.8, 0.5, 1.6] across octet, tetrakaidecahedron, and monolithic shields, respectively (Figure~\ref{fig:DimAnalysis}d-f). 
The exponent $\alpha$ governing initial velocity scaling $v_f^2 \propto v_i^{2 \alpha}$ closely matches that from prior particle impact experiments across all architectures examined \cite{Butruille2024}. However, the exponent $\beta$ governing the length scaling relation $v_f^2 \propto (r/h)^{\beta}$ varies significantly across architectures. Physically, decreasing $r/h$ with a larger $\beta$ leads to more rapid decreases in the final particle velocity. 
As $\beta_{\mathrm{mono}}>\beta_{\mathrm{octet}}>\beta_{\mathrm{tetrakai}}$, increasing monolithic or octet shield thickness should lead to greater enhancement of impact resistance than equivalent increases in tetrakaidecahedron shields. Interestingly, analysis of energy dissipation components across architectures have shown that octet and monolithic architectures outperform the tetrakaidecahedron in impact-boundary driven dissipation mechanisms such as shear and radial fracture \cite{Butruille2024}. Decreasing $r/h$ therefore implies a decrease in material participating in energy dissipation at or around the crater boundary relative to material compacting under the impacting particle.

This difference in relative efficiency in mitigating impact as $r/h$ varies across monolithic and architected material systems can inform the design of 3D laminated or heterostructured lattices. As material recruitment occurs along the impactor trajectory, the effective $r/h$ ratio evolves dynamically during impact, providing opportunities to optimize on a heterostructure design paradigm towards the design of next-generation impact-protection materials. 

\section{Programable cratering in heterostructured shields}
Composite and layered structures have seen extensive use in protective applications across centuries. Armor systems in antiquity featured layered systems of plates, chain, and cloth that served to prevent penetration across layers and disperse impact across a broader area of the armor \cite{Ramdayal2016}. Modern protective equipment such as sport helmets combine hard plastics and soft foams to achieve similar goals. In nature, characterization of biological protective materials such as the dactyl club of a mantis shrimp reveal stark differences in microstructure and function through the depth of the club. At the club surface, nanoparticle agglomeration serves as a preliminary protective mechanism. Underneath, the helical Bouligand structure provides labyrinthine crack deflection and dissipates energy through the creation of fracture surfaces \cite{Suksangpanya2017, Behera2021, Alderete2025}. In a similar manner to these composite systems, our heterostructure design paradigm leverages relative advantages of architected and monolithic layers to influence energy dissipation and crater formation during  ultra-high-speed particle impact.

To create our suspended heterostructured shields, we fabricated combinations of octet and monolithic layers upon our support structures (and base layer), to a total of six layers for all shields printed. As the impacting particle traverses through these heterostructures, the effective $r/h$ ratio and the relative impact-mitigating efficiency of each layer changes. To maximize the efficiency of these shields, we chose the octet (as opposed to the tetrakaidecahedron) for its superior mass-normalized impact mitigating properties \cite{Butruille2024}. 

In designing our heterostructured shields, the ratio and the ordering of octet and monolithic layers was varied with respect to the particle impact direction for a total of five designs, each demonstrating the importance of layer ordering and ratio on their impact response and allowing contrasts to single-architecture shields. For each octet layer being one unit cell thick, mass-equivalent monolithic layers corresponded to a 1.9 \textmu{}m thickness. These designs included (i) four layers of octet with one monolithic layer in both configurations (4O-1M, 1M-4O), (ii) four monolithic layers with one octet layer in both configurations (4M-1O, 1O-4M), and (iii) alternating monolithic and octet layers (1M-1O-1M-2O) as depicted in Figure~\ref{fig:DesignParadigm}a. For consistency, all designs list the layers in the order of impact. 
\vspace{10pt}

%%% Figure 3 %%%%%%%%%%%%%%%%%%%%%%%%%%%%
\begin{figure*}[t]
    \centering
    \includegraphics[width=12 cm]{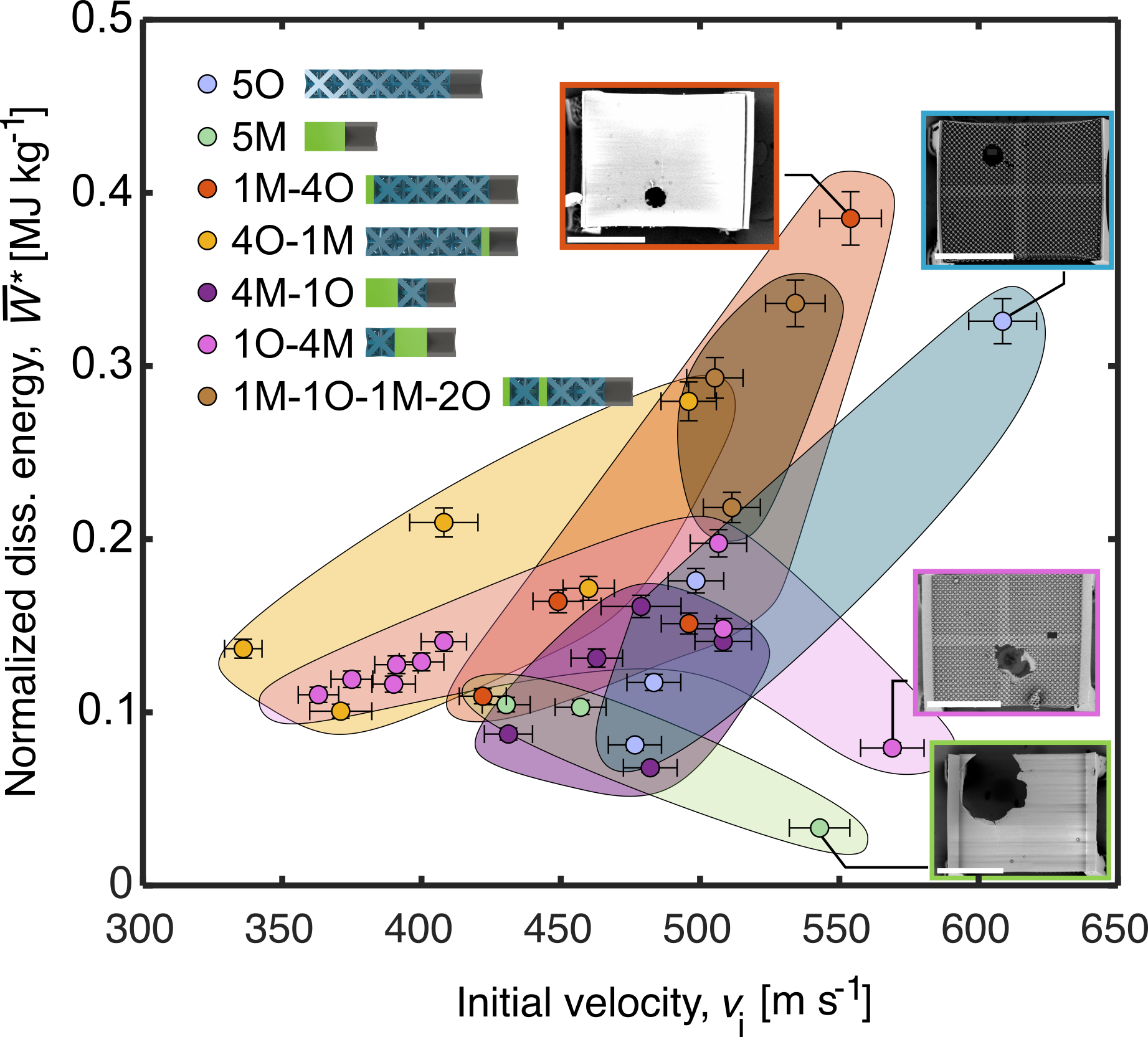}
    \caption{Incident particle velocity vs crater-mass-normalized energy dissipation for all suspended shield architectures, assuming craters with constant area through thickness. Scale bars, 100 \textmu{}m.}
    \label{fig:Ediss}
\end{figure*}
%%%%%%%%%%%%%%%%%%%%%%%%%%%%%%%%%%%%%%%%%%
\subsection{Post-impact tomographic crater reconstruction}
We performed microparticle impact experiments on all heterostructured designs with initial impact velocities $v_i$ ranging from 300 to 650 m s\textsuperscript{-1} (Figure S4) while also capturing the final impact velocity $v_f$ from each experiment. The single-architecture octet and monolithic shields behaved similarly to prior experiments, with well-localized craters in the octet shields contrasting with the catastrophic failure in the monolithic shields. This catastrophic failure was not observed in the heterostructured samples, with 1M-4O, 4O-1M, and 1M-1O-1M-2O shields exhibiting cratering behavior more similar to the octet architecture at the first layer. Quantitative comparison was performed through a crater-mass normalized energy dissipation metric $\bar{W}^*=\frac{1}{2}m_{\mathrm{sphere}}(v_i^2-v_f^2)/m_{\mathrm{crater}}$, where the particle mass was defined as $m_{\mathrm{sphere}}=\frac{4}{3}\rho_{\mathrm{sphere}}\pi/r^3$ and the crater mass as $m_{\mathrm{crater}}=\rho\bar\rho A_{\mathrm{crater}}h$. The crater area $A_{\mathrm{crater}}$, was assumed to be constant through-thickness, and was measured from top-down SEM images of the post-impact suspended shields. As shown in \textbf{Figure~\ref{fig:Ediss}}, the well-localized nature of craters in heterostructures 1M-4O, 1O-4M, and 1M-1O-1M-2O allowed for outperformance in the $\bar W^*$ metric when compared to both single-architecture suspended shields (5O, 5M), likely due to a combination of deformation mechanisms during impact and the crack-pinning properties of the octet lattice. 
\vspace{10pt}

%%% Figure 4 %%%%%%%%%%%%%%%%%%%%%%%%%%%%
\begin{figure*}[!t]
    \centering
    \includegraphics[width= 16 cm]{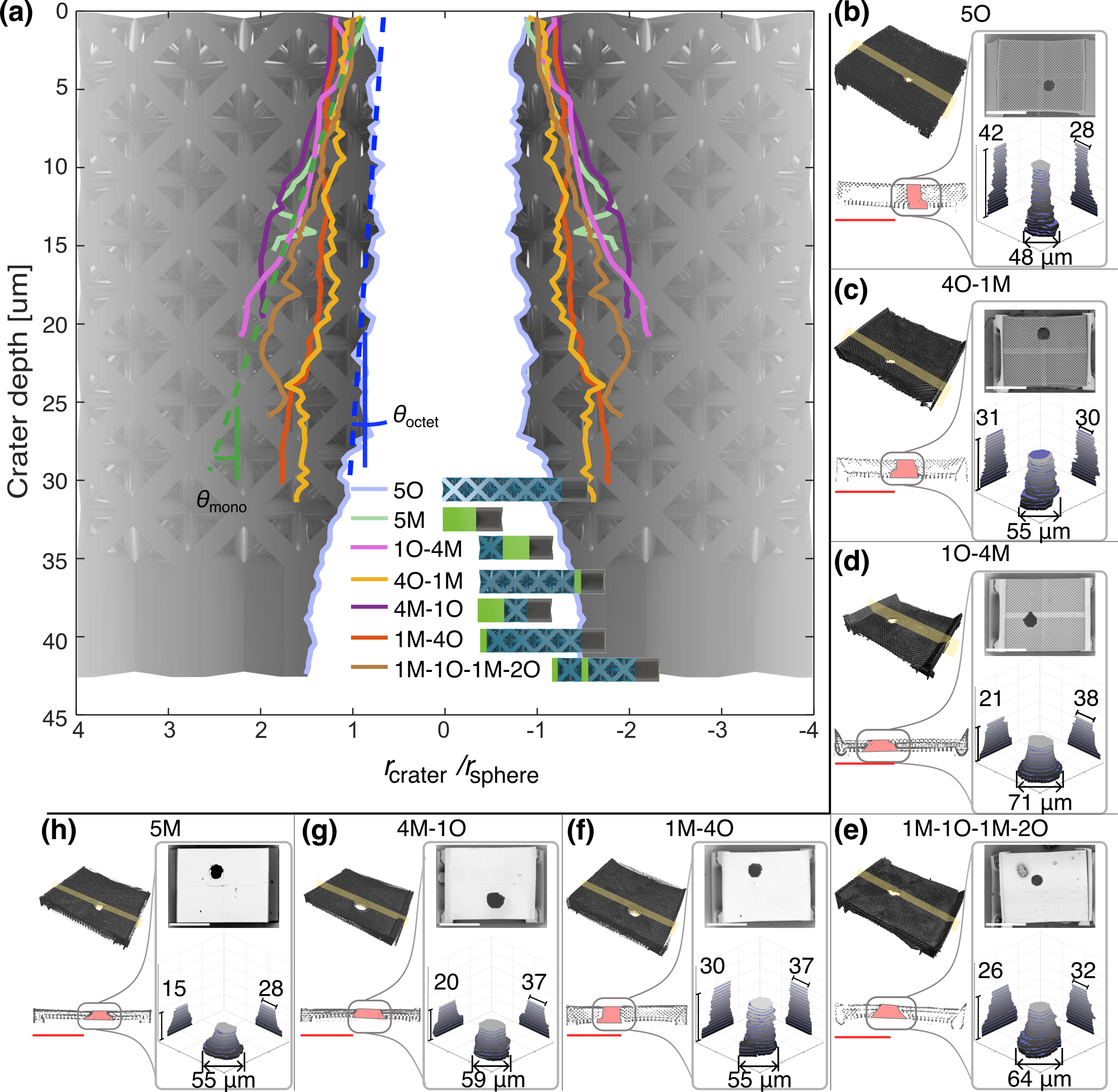}
    \caption{\textbf{a)} Crater profiles for tested shields, presenting crater radii normalized by the particle radius. XCT-enabled 3D reconstruction of representative post-impact suspended shields for \textbf{b)}5O, \textbf{c)} 4O-1M, \textbf{d)} 1O-4M, \textbf{e)} 1M-1O-1M-2O, \textbf{f)} 1M-4O, \textbf{g)} 4M-1O, and \textbf{h)} 5M architectures, respectively---including a cross-section view of the crater along with a top-down SEM micrograph and a 3D representation of the crater volume. Scale bars, 100 \textmu{}m.}
    \label{fig:Cratering}
\end{figure*}
%%%%%%%%%%%%%%%%%%%%%%%%%%%%%%%%%%%%%%%%%%
To examine the source of this superior performance in heterostructures, we sought additional information from the post-impact crater morphologies towards verifying our assumptions of a vertical-wall crater. To that end, we performed x-ray computed tomography (XCT) on the suspended lattices (XRM Versa 620, Experiments and Methods), reconstructing crater morphologies in single-architecture and heterostructured shields after impact with sub-micron resolution. 
From the XCT reconstructions, binary images corresponding to single-voxel-thick slices (thickness $h_k\approx 400$ nm) were used to quantify the evolution of the crater as a function of depth. For each of these slice images, we computed a relative density $\bar\rho_k$ based on the solid fill-fraction in that slice, and additionally defined a crater boundary. The stack of crater boundaries, each defining a crater with area $A_k$, was summed to provide the crater volume for each sample. At each slice, an effective crater radius $r_{\mathrm{crater}}=\sqrt{A_k/\pi}$ was calculated and compared to the incident particle radius and the depth of the crater from the shield surface. From these metrics, we confirmed the octet lattice crater to be the closest to a cylindrical crater, with the smallest deflection angle in the $r_{\mathrm{crater}}$ profile from the particle boundary (\textbf{Figure~\ref{fig:Cratering}}a-b). This angle was calculated from the slope of the line of best fit and found to be $\theta_{\mathrm{octet}}=11\degree$. In contrast, the monolithic shield crater showed the largest deflection from the particle boundary with $\theta_{\mathrm{mono}}=40\degree$, implying that the straight-wall assumption for the crater results is an underestimate of the crater mass (Figure~\ref{fig:Cratering}a,h).  
\vspace{10pt}

Following the trends for octet or monolithic-only shields, the smallest deflection of crater radii was observed for heterostructed shields including four octet layers (Figure~\ref{fig:Cratering}c,f), while shields with four monolithic layers had crater deflection profiles that matched or exceeded the deflection in the monolithic profile (Figure~\ref{fig:Cratering}d,g). In both cases, the shields featuring their monolithic layers closest to the shield surface had crater boundaries furthest from the particle boundary, resulting in increased mass utilization. The alternating heterostructured shield 1M-1O-1M-2O exhibited a deflection angle of $37\degree$, roughly equivalent to the angle deflections for the monolithic and 4M-1O shields  (Figure~\ref{fig:Cratering}e,g). This evidence indicates that the use of the heterostructure design paradigm not only suppresses catastrophic failure of the shields through avoiding radial crack propagation, but it also allows for control of crater formation, thus amplifying the amount of material that participates in an impact event. 
\vspace{10pt}
%%% Figure 5 %%%%%%%%%%%%%%%%%%%%%%%%%%%%
\begin{figure*}[t]
    \centering
    \includegraphics[width=16 cm]{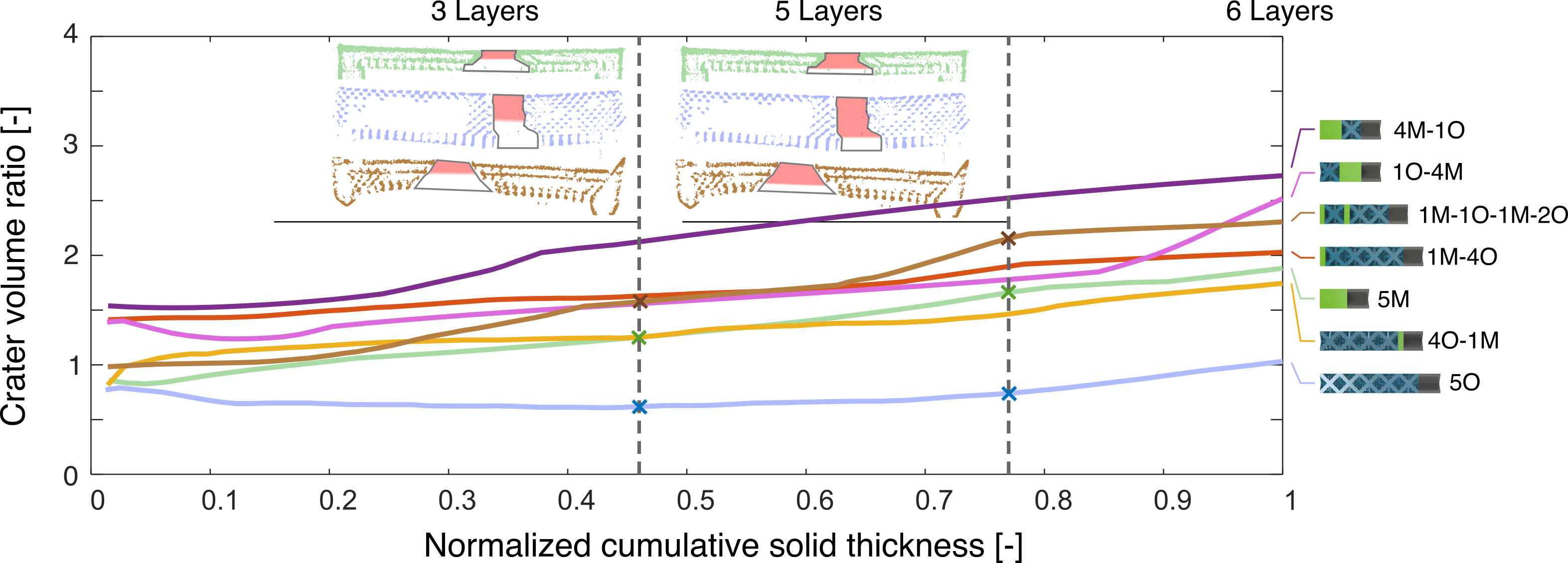}
    \caption{Normalized equivalent thickness versus normalized equivalent monolithic volume for all suspended shield architectures imaged with XCT. Cross-sections of select heterostructures at the particle impact crater location depicted at three layers and six layers of crater depth.}
    \label{fig:CumulativeVolume}
\end{figure*}
%%%%%%%%%%%%%%%%%%%%%%%%%%%%%%%%%%%%%%%%%%

To quantify the relation between crater morphology and material utilization in the shields upon impact, we computed the crater volume ratio $\bar V_n$ that was excavated from the crater as a function of $n$, the XCT shield slice index. This $\bar V_n$ metric was computed by summing the volume excavated to the $n$-th XCT slice, normalizing by the cylindrical volume directly under the impacting particle at that slice, and is then plotted against the normalized cumulative solid thickness $\bar h_{n}$ over $N$ total slices:

\begin{equation}
\label{eq:NormSum}
\begin{split}
    \bar V_{n}=\frac{1}{n}\sum_{k=1}^n  \frac{\bar\rho_k A_k h_k}{\bar\rho_k \pi r_{\mathrm{sphere}}^2 h_k},\\
    \bar h_{n}=\frac{\sum_{k=1}^n \bar\rho_k h_k}{\sum_{k=1}^N \bar\rho_k h_k}~.
\end{split}
\end{equation}

This metric demonstrated differences in mass utilization across shields (\textbf{Figure~\ref{fig:CumulativeVolume}}), highlighting the lowest mass utilization in the 5O shield while the 4M-1O architecture achieved the highest, with a relative difference of $\sim$240\% between them through the first five shield layers. Notably, this metric evidenced order-based mass recruitment trends, where 4M-1O and 1M-4O shields increased in $\bar V_n$ by 42\% and 29\% compared to their 1O-4M and 4O-1M counterparts at their five-layer thresholds, respectively. The 1M-1O-1M-2O shield also nearly matched the crater volume ratio of the 4M-1O architecture despite having only half the monolithic layers, indicating that these trends are not a simple function of the architecture ratio alone. To better understand these deflection trends, we turned to stress-wave propagation analysis---with corresponding finite element simulations across select shield architectures---towards gaining predictive capabilities of heterostructure crater profiles.

\subsection{Stress propagation analysis during cratering}
 As demonstrated in other materials systems, impact delocalization or the recruitment of additional mass laterally is enhanced by high in-plane wave speed of the material \cite{Hyon2018}. Experiments on multilayer graphene exhibited delocalized fracture and deformation, enabled by the high Young's modulus of the material \cite{Lee2014}. Beyond conducting experiments at ultra-high strain rates, we sought to obtain predictive measures for failure delocalization and thus energy dissipation through understanding of stress-wave propagation characteristics through a suspended plate. While damage is ultimately the result of a plastic wave (or compaction front) propagating through the material, the elastic precursor wave generated by the particle at the impact site can serve as a proxy for the emergence of delocalization, as demonstrated by analysis of graphene films displaying highly delocalized fracture behavior \cite{Lee2014} and finite element simulations of helical Bouligand architectures showing a correlation between stress wave delocalization during impact and macroscale cracking \cite{Meng2025}. Since direct observation or measurement of stress waves under our experimental conditions is challenging, we employ computational models to calculate elastic and plastic wavefront evolution over time, towards predicting stress and fracture delocalization in heterostructured suspended shields.

\begin{figure*}[t]
    \centering
    \includegraphics[width=14 cm]{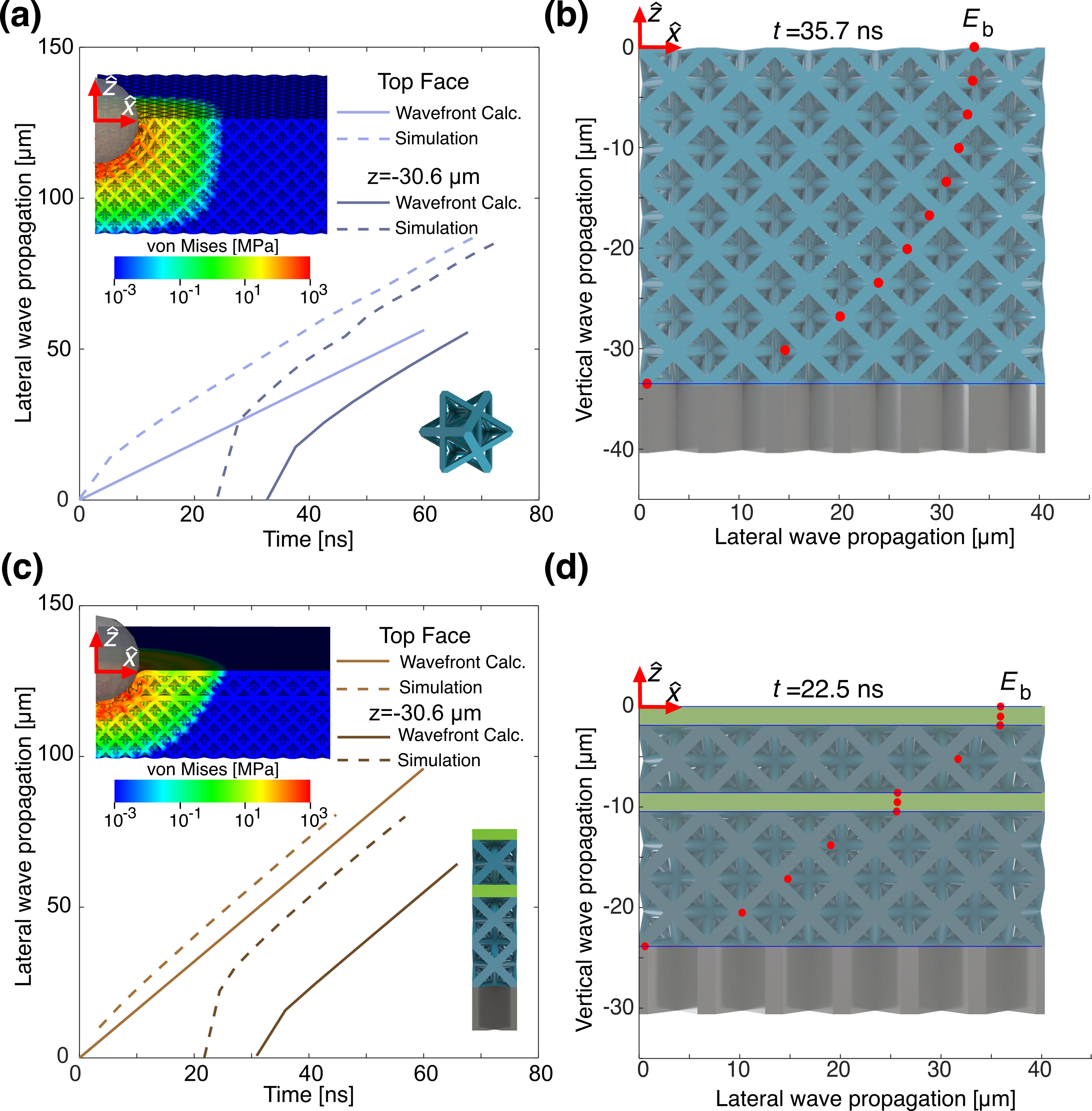}
    \caption{\textbf{a)} Comparison of lateral propagation of the elastic wavefronts in a finite element simulation using quarter-symmetry of an octet shield, depicting propagation over time with an elastic wave speed of $c_\mathrm{{octet}}$= 939 m s\textsuperscript{-1} (Experiments and Methods, Eq. \ref{eq:2LWavefront}). Propagation distance is reported relative to the center of the particle, while time is relative to initial particle contact. \textbf{b)} The calculated octet elastic wavefront at time \textit{t}=35.7 ns, which is when the elastic wavefront reaches the bottom face of the final octet layer. \textbf{c-d)} Comparison of lateral elastic stress propagation in a quarter-symmetry 1M-1O-1M-2O suspended shield, and the calculated lateral elastic stress propagation over time, with a respective snapshot of the calculated wavefront at time $t$=22.5 ns.}
    \label{fig:SimWavefront}
\end{figure*}

We performed finite element simulations of the LIPIT experiments on two structures, 5O and 1M-1O-1M-2O, to identify differences in the stress wave propagation as a function of the layer architecture of these shields. Using tracer nodes at the top surface and a depth of 30.6 \textmu{}m (Experiments and Methods)---corresponding to the thickness of the 1M-1O-1M-2O shield---we tracked the evolution of the elastic wavefront in both structures (\textbf{Figure~\ref{fig:SimWavefront}}a,c). The results indicate a faster in-plane wave speed for the 1M-1O-1M-2O structure, both on the top layer and bottom layer, indicating an effective stress-distribution functionality of the first monolithic layer. Frames of the von Mises stress field from these simulations, shown as insets (Figure~\ref{fig:SimWavefront}a,c), confirms that the elastic wave profile is extended laterally in the alternating architecture shield. While this analysis provides insight towards understanding the mechanisms of wave propagation in selected designs, exploring the vast design space of heterostructured architected materials with explicit nonlinear finite element simulations is computationally intractable. Therefore, we proceed with a simplified wave propagation model that may prove useful for evaluating and developing impact-mitigating architected shield designs.

 %%% Figure 6 %%%%%%%%%%%%%%%%%%%%%%%%%%%%
%%%%%%%%%%%%%%%%%%%%%%%%%%%%%%%%%%%%%%%%%%

This simplified estimation of stress wave propagation through a heterostructure was performed with the assumption that the stress wave travels isotropically within a given layer and at constant speed. At time $t$ after particle contact, taking a constant wave speed $c_p$ through a given architecture, a wave can propagate a distance $t c_p$. The lateral distance $x$ away from the initial particle contact point can then be calculated as $x=\sqrt{(t c_p)^2-z^2}$ where $z$ is the depth. At constant time $t$ and varying $z$ for a single architecture, taking $c_p=c_{\mathrm{octet}}$, where $c_\mathrm{{octet}}=939$ m s\textsuperscript{-1} is the average elastic wave speed for the octet in the in-plane direction \cite{Butruille2024}, an elastic wavefront boundary $E_b$ at time $t$ can be calculated and compared to that obtained from the finite element models (Experiments and Methods, Figure~\ref{fig:SimWavefront}a,b). For heterostructured shields where $c_p$ varies with architecture in each layer, neglecting any coupling effects from layer boundaries, the fastest path for wave propagation is rarely a straight line beyond the first layer. We solved for the wavefront in a two-layered heterostructure by optimizing the path that the stress wave travels given $z$ and $t$. An analytical solution for the two-layer heterostructure case is tractable, but we adopted a numerical approach in varying $x$-distance traveled in each material layer and choosing the path that propagates the wave furthest in $x$ given the constraints $t$ and $z$ (Experiments and Methods). To evaluate the quality of this method, we compared it to stress propagation in the finite element models at two constant depths---the top face depth of zero and a second depth of 30.6 \textmu{}m, corresponding to the bottom face of the 1M-1O-1M-2O shield---and measured the average time at which the stress wavefront reaches a particular lateral displacement between the $\langle$100$\rangle$ and  $\langle$110$\rangle$ orientations. At those depths, time was varied and maximum lateral displacement calculated given average longitudinal elastic wave speeds $c_{\mathrm{octet}}$ and $c_{\mathrm{mono}}$. This wavefront calculation method provided good qualitative agreement with the computational models when fixing either depth or propagation time (Experiments and Methods, Figure~\ref{fig:SimWavefront}b, d). The lateral propagation speed $\partial x/\partial t$ at large $t$ also showed good agreement across both simulated shields (Figure~\ref{fig:SimWavefront}a,b). At small $t$, increased lateral propagation in the simulations could be attributed to material deformation occurring near the particle at speeds still significant compared to the elastic wave speed. 
 %%% Figure 7 %%%%%%%%%%%%%%%%%%%%%%%%%%%%
\begin{figure*}[t]
    \centering
    \includegraphics[width=14 cm]{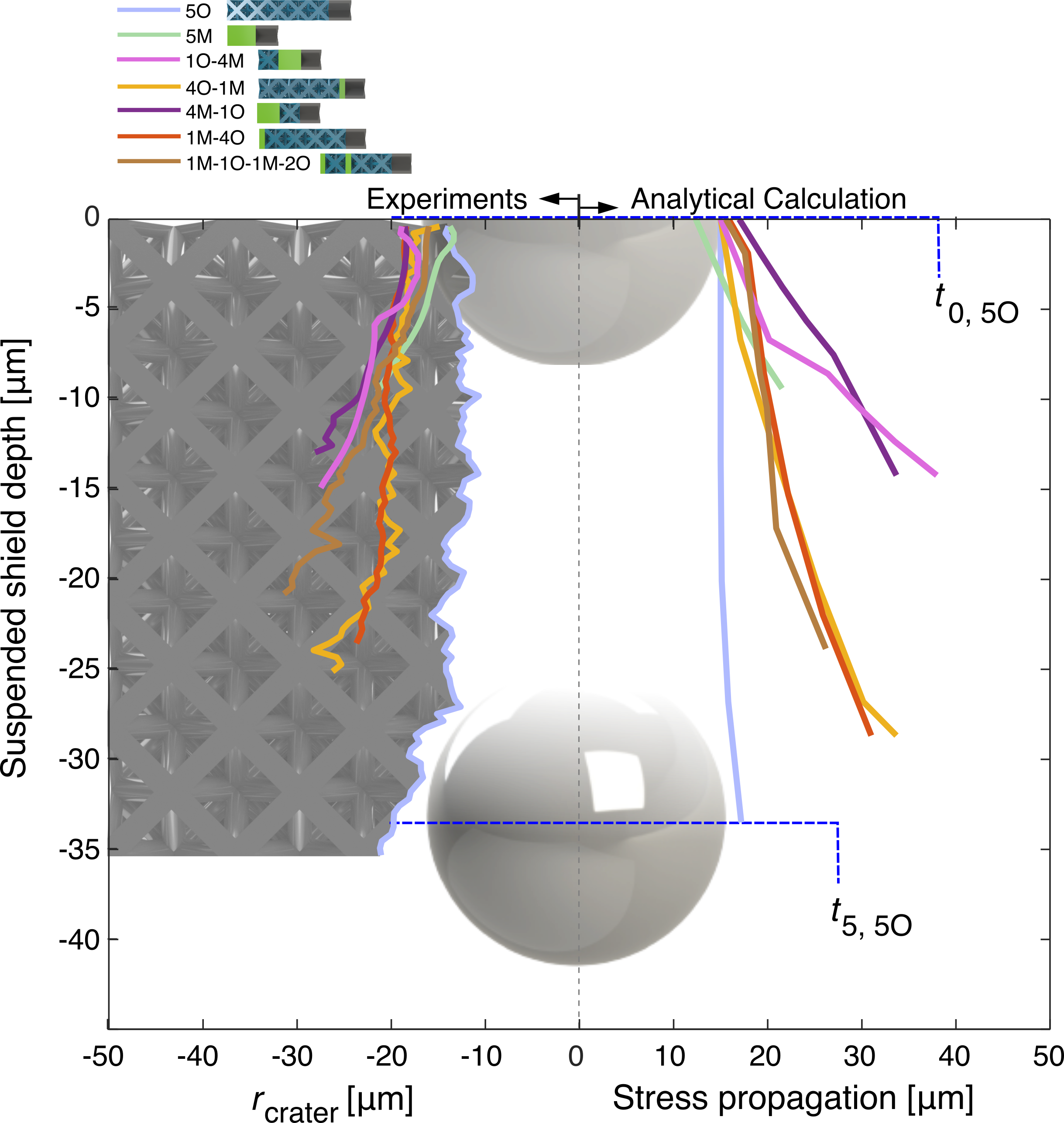}
    \caption{Comparison of $r_{\mathrm{crater}}$ and plastic wavefront propagation through all suspended shields to a depth of five layers. A particle with shield-specific $v_i$ and $v_f$ (Table S1) linearly interpolated between layers is used to calculate $t_k$, the time from particle contact with the top layer to when the particle side-edge passes a particular layer boundary. The top boundary has this time denoted with $t_0$, and the bottom boundary $t_5$. This time is used to calculate plastic wavefront propagation in the in-plane direction together with best-fit isotropic plastic wave speeds in octet and monolithic layers respectively.}
    \label{fig:Wavefront}
\end{figure*}
%%%%%%%%%%%%%%%%%%%%%%%%%%%%%%%%%%%%%%%%%%

Although determining the exact fracture criterion of our constituent material at these ultra-high strain rates is intractable with our current methods, we turn to plastic-stress-wave propagation across our suspended shields as a proxy for qualitative similarity between stress wavefront boundaries and the resulting crater boundaries. Assuming this direct correlation between the arrival of a plastic stress wave and localized failure at that site, an estimate of this plastic wave speed $c_p$ was obtained by fitting the experimentally determined crater profiles. We confirmed physical validity of these speeds by computing $c_p$ using experimental stress-strain curves of octet and monolithic samples, where the analytically determined solution for $c_p$ in both shields was at or above the minimum possible experimental values (Supporting Information, Figure S5).

To provide a first-order estimate for this plastic wave speed, we first calculate the time $t_k$ the stress wave has to travel at each layer $k$ as $t_k=\sum_{j=0}^{k}v_j h_j$ where $v_j$ and $h_j$ are particle velocity and thickness in individual layers. This time $t_k$ is the cumulative time it takes for the center of a particle to cross the $k$-layer interface, with initial and final velocities corresponding to the measured values for each  shield (Table S1). Here, we let $v_i$ decrease linearly to $v_f$ regardless of layer architecture.  Note that $h_0$ is taken to be 15 \textmu{}m, the nominal radius of an impacting particle and $v_0=v_i$. As the stress wave propagating outwards from the particle is traveling through deformed material, we assume it travels at the plastic wave speed in the octet and monolithic architectures, with $c_{p,~\textrm{5O}}=$ 370 m s\textsuperscript{-1} and $c_{p,~\textrm{5M}}=$ 524 m s\textsuperscript{-1}. To partially account for how $c_p$ may vary through the impact event, the minimum wavefront boundary at each layer is set at the particle radius. With the plastic wave speed and propagation time known in each heterostructure layer, we numerically solve for plastic stress propagation boundary in each heterostructure by parametrically varying the path the plastic wave travels from initial particle contact to boundary layer $h_k$ in time $t_k$, optimizing for the lateral distance $x_k$ from the initial particle contact point (\textbf{Figure~\ref{fig:Wavefront}}). 

As depicted in Figure~\ref{fig:Wavefront}, the calculated stress propagation profiles show strong qualitative agreement with their respective experimental crater profiles. Heterostructured shields with a greater number of monolithic layers had broader stress propagation profiles, while suspended shields with majority octet layers had stress propagation boundaries that began approaching cylinders. Furthermore, heterostructures with prior monolithic layers had greater lateral propagation than their order-switched counterparts in early layers, mirroring the experimental cratering profile trends discovered through XCT (Figure~\ref{fig:Wavefront}, left). However, the simplicity in our analysis failed to consider how interlayer coupling changed the relation between lateral stress propagation and the crater profile boundary. Qualitatively, the alternating heterostructure cratering profile was broader than its stress propagation boundary, suggesting additional  broadening mechanisms other than the lateral speed at which stress propagated, especially in later layers. Additionally, this analysis falls short in predicting how differences in $v_i$ between single-architecture and heterostructured shields may also alter the failure-associated plastic wave speed $c_p$ from that originally fit. Both 1O-4M and 4O-1M had initial particle velocities below their order-switched counterparts, increasing the time window for stress propagation (Table S1). Ultimately, calculating stress propagation boundaries through heterostructured suspended shields provides insight into how stress propagation speeds across architected layers drive dynamic cratering formation for impact mitigation applications. Through these calculations, expectations can be formed for efficient and absolute energy dissipation in heterostructured or laminated shields subject to projectile impact.

\section{Conclusion}
In the search for lightweight impact-mitigating materials, single-architecture suspended lattice shields with an octet lattice microstructure have demonstrated their ability to efficiently dissipate energy through the formation of a nearly cylindrical compaction shock localized to the impacting particle, outperforming other architectures. Compared to monolithic suspended shields of the same constituent material---which are stiffer but allow for stochastic radial fracture to occur during impact---these lattice architectures have localized deformation leading to more efficient energy dissipation on a mass-normalized basis \cite{Butruille2024}. In this work we presented heterostructures consisting of layers of 3D architected materials (octet lattices) with monolithic layers of the same bulk material, and have demonstrated their ability to shape impact mitigation efficiency by taking advantage of benefits in both monolithic and lattice architectures. In our heterostructured suspended shields, the crack localization and compaction response of the octet lattice was combined with the stress delocalization of the monolithic layers, shaping impact events in ways similar to the 3D microstructure gradation in biological systems such as the dactyl club of a mantis shrimp \cite{Behera2021, Alderete2025}. This limited catastrophic failure and improved mass-normalized energy dissipation efficiency by more than 50$\%$ compared to single-architecture lattices across the impact velocities we explored (Figure~\ref{fig:Ediss}).

We investigated this improvement in energy dissipation response from two directions, the first being dimensional analysis. By studying the effect of varying thickness in suspended shields, we determined a relation between the final particle velocity $v_f$, the particle radius $r$, and the shield thickness $h$ in each architecture, resulting in a form $v_f^2\propto\left(\frac{r}{h}\right)^\beta$. This exponent $\beta$ varied between architectures with $\beta_{\mathrm{octet}}=0.8$ and $\beta_{\mathrm{mono}}=1.6$, indicating which architecture was relatively more effective at lowering $v_f$ as $r/h$ increased or decreased. The heterostructure framework used these relative advantages between individual suspended architecture layers with the understanding that $r/h$ evolved through the course of particle impact, with $r$ effectively increasing beyond the radius of the impacting particle. This recruitment of material laterally during an impact event was linked to the crater boundary profiles for heterostructured suspended shields, which led to our second method of analyzing particle impact in these shields through plastic stress propagation. The speed of plastic stress propagation and the resulting stress propagation profile were fit to x-ray tomographs of the post-impact crater profiles in the octet and monolithic suspended shields, allowing us to calculate stress-propagation profiles for heterostructured shields and compare them to their respective experimental crater profiles. This agreement supported our hypothesis of the stiffer monolithic layers leading to increased lateral recruitment of material in our heterostructures, thus broadening our crater profiles and increasing mass recruitment without catastrophic failure (Figure~\ref{fig:Wavefront}). 

Relative efficiencies in different architectures at different stages in an impact event can work harmoniously in efficiently dissipating the energy of that impact. The design of these laminated architectures currently relies on simplified stress propagation assumptions through multiple architected layers, but future studies are necessary to characterize the effect of interlayer coupling and anisotropy in failure and stress propagation, ultimately relating these complex effects to energy dissipation and crater formation under particle impact. Heterostructured and functionally graded architectures are ultimately a promising avenue towards the design of materials for impact mitigation, with both challenges and opportunities ahead to create useful predictive metrics and also make these generalizable across material systems and length scales.

\section{Experiments and Methods}
\subsection{Suspended lattice fabrication}
Suspended lattice shields were fabricated using two-photon lithography (Nanoscribe PPGT2) with the 63x feature set and IP-Dip photoresist, with a laser scan speed of 10,000 \textmu{}m s\textsuperscript{-1}, and a laser power of 30 mW. Constituent material properties for this photoresist, at these printing parameters, were obtained from uniaxial compression of monolithic pillars. Two mirrored support structures were printed a distance of 195 \textmu{}m apart to a height of 300 \textmu{}m. Bloch wave analysis on these unit cells confirmed that six-layer single-architecture lattices are decoupled from the support structures during a characteristic impact event. A subsequent base layer consisting of a 28$\times$26$\times$1 tesselation of a projected cross with 1.2\textmu{}m-wide spokes and a nomial thickness of 7.5 \textmu{}m was printed on the support structures, 262.5 \textmu{}m above the substrate. In the single-architecture lattice shields, octet and tetrakaidecahedron architectures with a nominal unit cell of 7.5 \textmu{}m in size, corresponding to a relative density of 26\%, were fabricated on top of the base layer. Monolithic plates of different thicknesses were printed to ensure they were mass-equivalent to the single architecture samples. After printing, these samples are developed for 2 hours in propelyne glycol methyl ether acetate, then subjected to a 5 minute rinse in isopropyl alcohol followed by critical point drying (Tousimis 931). Following development-induced shrinkage, the final unit cell dimensions were determined to be  6.7$\pm$0.1 \textmu{}m and the slightly elliptical beams for the major and minor axes in the octet lattice were 1.5$\pm$0.3 and 1.12$\pm$0.05 \textmu{}m respectively, and 2.0$\pm$0.1 and 1.6$\pm$0.1 \textmu{}m for the tetrakaidecahedron. The as-fabricated relative density for these lattices was determined to be 33$\pm$1.2\%. Heterostructured suspended shields were printed and developed with the same parameters as the single-architecture shields.

\subsection{X-ray Computed Tomography Data Processing}
Using an XRM Versa 620, each sample was imaged with the 20x objective, multibinning of two voxels, and an exposure time of 3 seconds across 1600 total frames resulting in a total scan time of $\sim$2.5 hours for each shield. X-ray computed tomography data was first uploaded to Dragonfly for solid model extraction. The scan space was cropped to include only the suspended shield, and denoising was performed where each `island' of fewer than 600 connected voxels was deleted from the solid model. Cross-checking the processes scans with top-down SEM images of the shield allowed for manual verification. The 3D model was then split into binary images at each voxel plane through the lattice thickness. Crater boundary profiles at each slice were selected manually, with the encapsulated areas stacked vertically to build the crater volume. The average centroid for each crater slice area was taken to be the in-plane crater center for the purpose of plotting each volume. 

\subsection{Finite element (FE) analysis}
The geometries of the FE simulations were made to match the geometries reported in the as-printed samples, except the extruded-cross base layer which was replaced by an additional layer of octets to simplify the model. Quarter-symmetry boundary conditions were applied to reduce the domain to a 16$\times$13$\times$1 tessellation of columns, with each column having six monolithic or octet layers corresponding in order and proportion to the simulated heterostructure design. A 30 \textmu{}m diameter particle with initial velocity of 500 m s\textsuperscript{-1} was impacted onto the freestanding shield at $x=y=z=0$. Displacements were fixed on the $y$-$z$ surface furthest from impact. The 5O and 1M-1O-1M-2O shields were modeled with meshes consisting of $48\times 10^6$ and $54\times 10^6$ linear tetrahedral elements, respectively. Lagrangian tracers were distributed along the $x$ axis and along $x=y$ on the top surface ($z=0$), and at a depth equivalent to the bottom of the 1M-1O-1M-2O heterostructure ($z=-30.6$ \textmu{}m). The time at which the von Mises stress is nonzero at these tracer points was used to determine the wavefront of the elastic precursor in Figure~\ref{fig:SimWavefront}.

The material model and parameter calibration followed prior work\cite{Butruille2024}, and consisted of a power-law hardening model with a von Mises yield surface and ductile failure based on the tearing parameter $t_p$ proposed by Wellman \cite{Wellman2012}. As expected, we observed sensitivity in the plastic wave propagation and cratering profile when varying the plasticity model and failure criteria. With localized strain rates approaching $10^6$ s\textsuperscript{-1} and given the uncertainty in the plastic and failure responses of the constituent material at strain rates $\geq 10^3$ s\textsuperscript{-1}, the focus of the FE simulations was on capturing the elastic wave propagation. This was found to be insensitive to changes in the plasticity and failure models, as expected. The plasticity and failure models were only required to provide a reasonable collapse of the shield as the elastic wave and particle propagate through the thickness of the shield.

\subsection{Plastic wavefront calculation}
Stress-wave propagation through our suspended shields was used to judge lateral mass recruitment and predict cratering during impact. We began with the simplification that our stress propagation was isotropic in our bulk material and the octet lattice, and the assumption that stress propagation through monolithic-lattice interfaces occurred without impediment in speed. With stress propagation originating at some point on the uppermost layer, the lateral distance that the stress wave could travel at a given layer interface with depth $z$ over some time $t$ can be calculated. In the case of a single architecture (monolithic or octet) the boundary of stress propagation associated with some stress-wave propagation speed is hemispherical. When considering a heterostructure of two architectures with different stress-wave propagation speeds $c_1$ and $c_2$, an optimization problem must be solved to determine the maximum lateral distance stress can propagate into the second layer at time $t$ and depth $z\leq z_1$, where $z_1$ is the depth of the interface between architectures. This one-interface optimization problem has an analytical solution derived by optimizing the path the stress travels to depth $z$ for maximum lateral distance $x$. We begin this problem considering that for the stress to travel to $x_c$, it must do so in a sum of linear segments between its origin point, the architecture interface, and its end point. The key to solving this problem is thus picking some point $(x_1, z_1)$ on the architecture interface at depth $z_1$ where the final segment length satisfies the time constraint $t$ and maximizes $x$. Letting $x_2$ and $z_2$ be the lateral width and height of the second segment, and $t_1$ and $t_2$ be the time it takes for stress to propagate in the first and second segments, we have $t_1+t_2=t$, where $\sqrt{x_k^2+z_k^2}/c_k=t_k$ relates the travel path and the time it takes for stress to travel in each architecture. In the first layer, where $z\leq z_1$, the maximum lateral distance the stress can travel in time $t$ is simply $x=\sqrt{(c_1 t)^2-z^2}$. Where $z>z_1$,  we substitute $x_2=x-x_1$ into our expression for total time $t=\sum_{i=k}^2 \sqrt{x_k^2+z_k^2}/c_k$ to obtain

\begin{equation}
\label{eq:2LWavefront}
\begin{split}
    t=\frac{\sqrt{x_1^2+z_1^2}}{c_1}+\frac{\sqrt{(x-x_1)^2+z_2^2}}{c_2}.
\end{split}~
\end{equation}

We can maximize $x$ by reordering equation \ref{eq:2LWavefront} to express $x$ as a function of $x_1$, solving for $\frac{\partial x}{\partial x_1}=0$. In the case for our 1M-1O-1M-2O heterostructured shield where there are three architecture interfaces excluding that of the base layer, solving for the maximum $x$ at depth $z$ follows a similar process but is no longer analytically tractable for depths greater than the second architecture interface. Instead, points $(x_k,z_k)$ at interfaces are chosen to satisfy $t=\sum_{i=k}^n \sqrt{x_k^2+z_k^2}/c_k$ and varied parametrically across physically reasonable points $x_k<c_\mathrm{{max}}t$, where $c_{\mathrm{max}}$ is simply the highest stress propagation speed of any layer within the heterostructure. In practice, the bound on intermediate points $x_k$ can be further reduced by considering the time it would take to propagate strictly downwards from $(x_k,z_k)$ to $(x_k,z)$, reducing the computational burden for heterostructures with many layers.

\section*{Acknowledgements}
We acknowledge financial support from DEVCOM ARL Army Research Office through the MIT Institute for Soldier Nanotechnologies (ISN) under Cooperative Agreement number W911NF-23-2-0121. This work was supported in part by high-performance computer time and resources from the DoD High Performance Computing Modernization Program. This work was carried out in part through the use of ISN and MIT.nano facilities. We also acknowledge support from MIT.nano laboratory staff, especially to J. Weaver for assistance in XCT imaging.

\medskip
\textbf{Conflict of Interest} \par
The authors declare no conflict of interest.
% References
\bibliographystyle{unsrt}

\medskip

\newpage
\renewcommand{\thefigure}{S\arabic{figure}}
\setcounter{figure}{0}
\renewcommand{\thetable}{S\arabic{table}}
\setcounter{table}{0}
\renewcommand{\theequation}{S\arabic{equation}}
\setcounter{equation}{0}
\section*{Supporting Information}
\subsection*{Supporting figures}
Figure S1. Imaging of multilayered tetrakaidecahedron shields subjected to LIPIT.

Figure S2. Imaging of multilayered monolithic shields subjected to LIPIT.

Figure S3. Initial and final particle velocities of single-architecture multilayered shields.

Figure S4. Initial and final particle velocities of six-layer single-architecture and heterostructured shields.

Figure S5. Uniaxial compression behavior and corresponding wave speeds in monolithic IP-Dip pillars and octet lattices.
\subsection*{Supporting tables}

Table S1. Initial and final velocities across tomographically imaged shields.
\subsection*{Supporting text}

\paragraph{Impact in multilayered shields}
In addition to the multilayered octet shields shown in Figure 2a-c, tetrakaidecahedron shields (Figure S1) and monolithic shields (Figure S2) were printed with four to ten layers---including the base layer---and subjected to laser induced particle impact testing. The monolithic shields were multilayered in the sense they incorporated a monolithic plate of thickness 4.6$\pm$0.7 \textmu{}m, 7.6$\pm$1.2 \textmu{}m, 10.6$\pm$1.7 \textmu{}m, or 13.7$\pm$2.2 \textmu{}m, approximately mass-equivalent to four, six, eight, and ten lattice layers, respectively. 

\paragraph{Particle velocity measurements}
Initial and final velocities were measured from high-speed camera frames of impact events across single-architecture shields of varying layer count (Figure S3), or six-layer single-architecture and heterostructured shield designs (Figure S4). These velocities were used in conjunction with post-impact crater imaging for the calculation of mass-normalized energy dissipation metrics (Figure 3), and the estimation of lateral stress-wave propagation for comparison to crater boundaries (Figure 6a,c, Figure 7).

\paragraph{Plastic wave speed fit}
The plastic wave speed $c_p$ for the octet and monolithic shields was fit to their respective experimental crater profiles as shown in Figure 4a to a depth of five layers. A wavefront for a trial value of $c_p$ was calculated with initial and final particle velocities equal to those measured across XCT-imaged 5O and 5M shields (Table S1). Letting $c_p$ vary with an increment of 1 m s\textsuperscript{-1}, the wave speeds that minimized the root mean squared error between the trial wavefront and the crater profile for the octet and monolithic shields were $c_{p, \textrm{5O}}=370$ m s\textsuperscript{-1} and $c_{p, \textrm{5M}}=524$ m s\textsuperscript{-1}, respectively. As validation, these plastic wave speeds were compared to the \emph{approximated} plastic wave speeds associated with the stress-strain curves an octet lattice or monolithic pillar under uniaxial compaction (Figure S5). We note that these approximated plastic wave speeds serve as a mere verification, since the experiment was performed at strain rates that are far from those in the impact experiments. Here, the `true' approximated plastic wave speed to large strain is computed as $c_{p, ~\textrm{mono}}=\sqrt{\frac{\partial\sigma_t/\partial\epsilon_t}{\rho}}$ for the monolithic polymer and $c_{p,~\textrm{octet}}=\sqrt{\frac{\partial\sigma/\partial\epsilon}{\rho_t}}$ for the octet, where $\sigma_t=\sigma(1+\epsilon)$ is the true stress, $\epsilon_t=\textrm{ln}(1+\epsilon)$ is the true strain, and $\rho_t=\rho/(1+\epsilon)$ is the true density. Note that the strain $\epsilon$ in these equations is, by convention, negative when compressive. These substitutions result from a simplifying assumption of plastic incompressibility in the monolithic polymer, and the simplification of affine compressibility in the porous octet architecture. Substituting these values into the plastic wave speed equations and solving in terms of $\sigma$, $\epsilon$, and $\rho$ results in

\begin{equation}
\label{eq:PlasticWavespeed}
\begin{split}
    c_{p,~\textrm{mono}}=\sqrt{\frac{(1+\epsilon)^2}{\rho} \frac{\partial\sigma}{\partial\epsilon}+\frac{(1+\epsilon)}{\rho}\sigma},\\
    c_{p,~\textrm{octet}}=\sqrt{\frac{(1+\epsilon)}{\rho}\frac{\partial\sigma}{\partial\epsilon}}.
\end{split}~
\end{equation}

In the small strain limit, both equations converge to $c_p=\sqrt{\frac{\partial\sigma/\partial\epsilon}{\rho}}$. Comparing the analytically fit plastic wave speeds $c_{p,~\textrm{5O}}$ and $c_{p,~\textrm{5M}}$ for octet and monolithic shields to the experimental approximations in Figure S5 places them below the elastic wave speeds as expected, with the monolithic one consistent with the wave speed at $\epsilon \approx 0.1$, while the octet's was at $\epsilon \approx 0.4$. 
% \newpage

\begin{figure}[h]
    \centering
    \includegraphics[width=\textwidth]{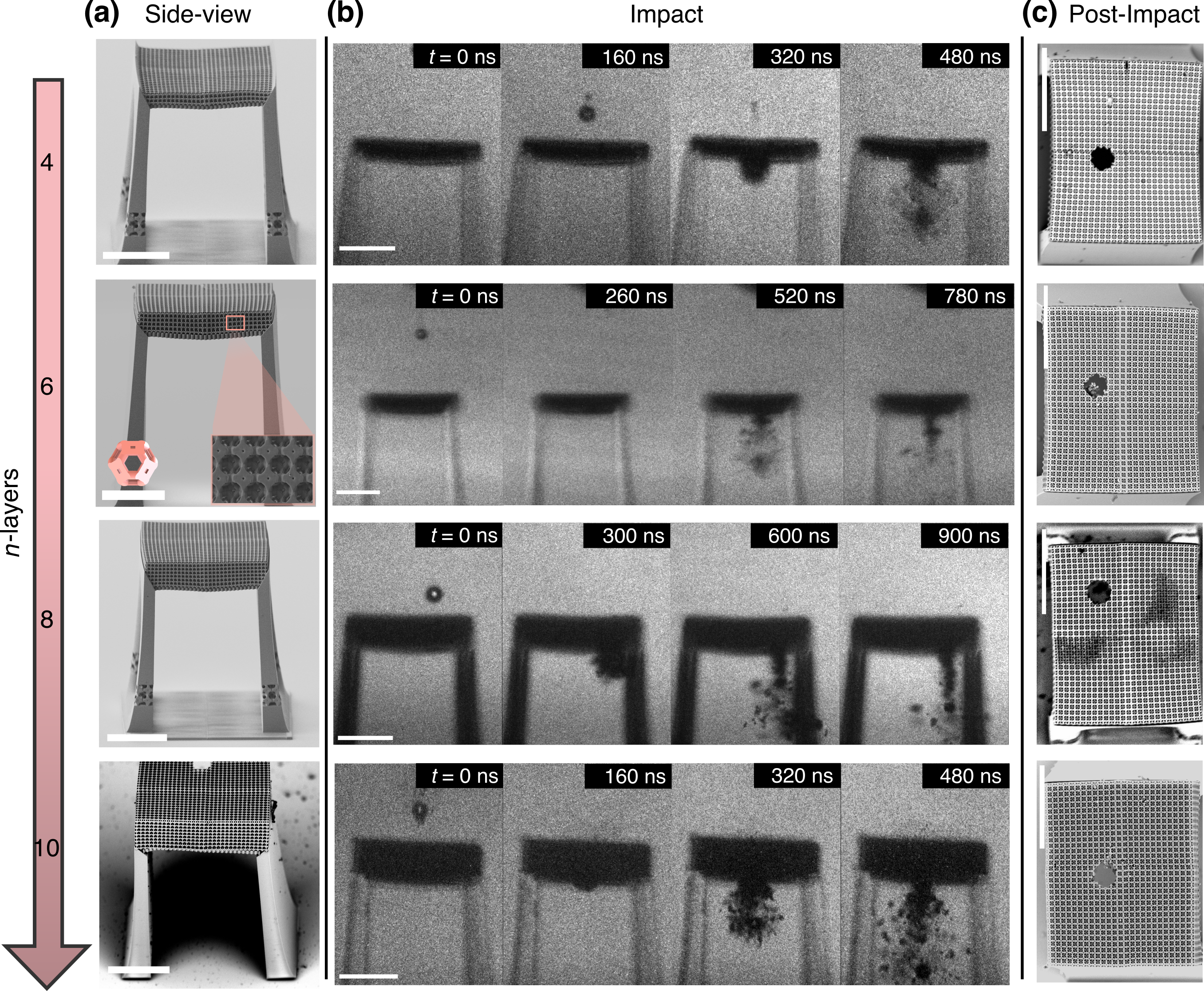}
    \caption{\textbf{a)} Multilayered pre-impact tetrakaidecahedron suspended lattices and \textbf{b)} high speed particle impact imaging at four, six, eight, and ten unit cell thickness (including the base layer), with impact velocities of 704, 573, 559, 912 m s\textsuperscript{-1}, respectively. \textbf{c)} Post-impact SEM of tetrakaidecahedron lattice craters across shield thicknesses.}
    \label{fig:Tet_SEMSIM}
\end{figure}

\begin{figure*}[t]
    \centering
    \includegraphics[width=\textwidth]{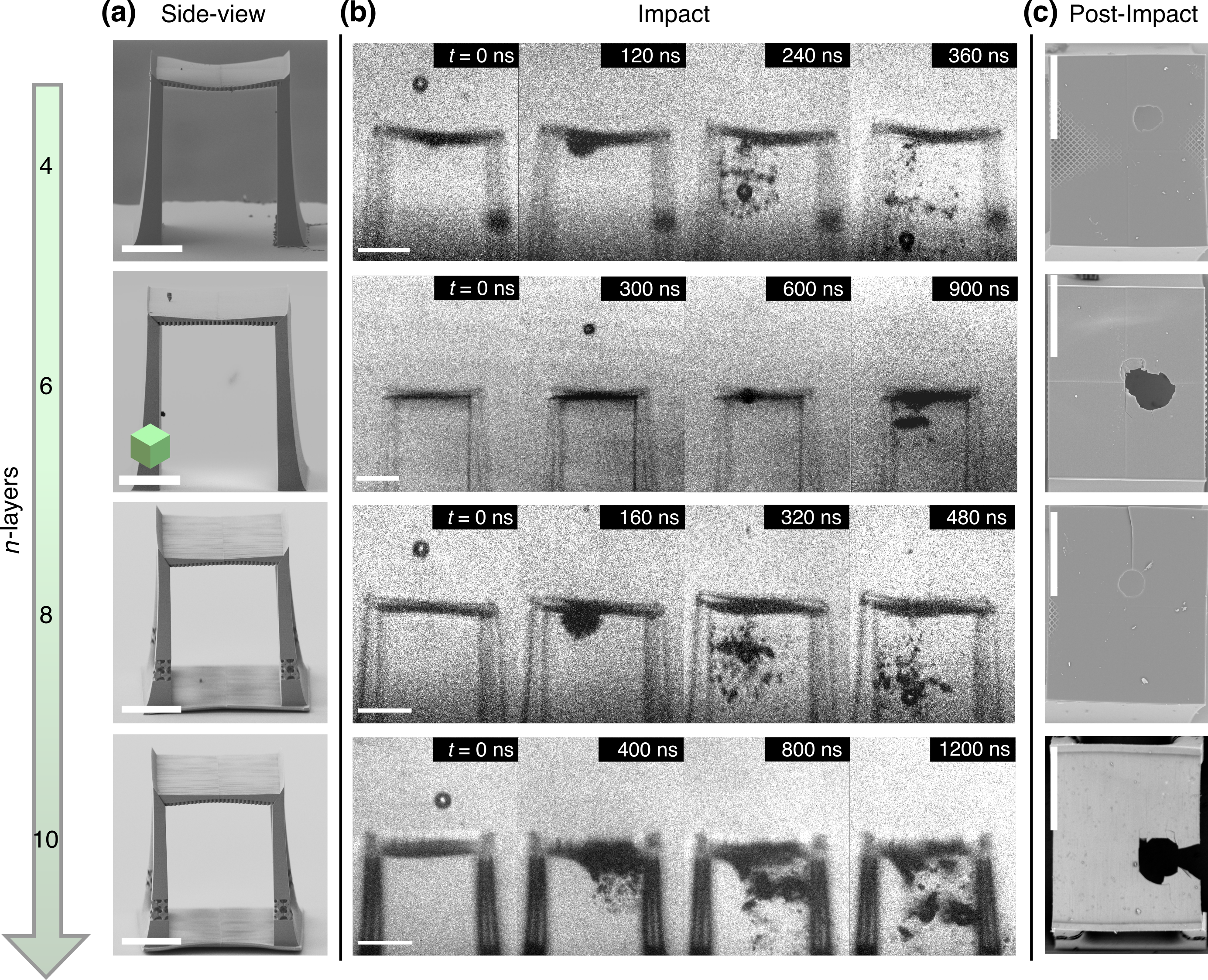}
    \caption{\textbf{a)} Pre-impact suspended monolithic shields and \textbf{b)} high-speed particle impact imaging at shield thicknesses roughly mass-equivalent to four, six, eight, and ten unit cells (including the base layer), with impact velocities of 907, 473, 408, and 670 m s\textsuperscript{-1}, respectively. \textbf{c)} Post-impact SEM of tetrakaidecahedron lattice craters across shield thicknesses}
    \label{fig:Mono_SEMSIM}
\end{figure*}

\begin{figure*}[t]
    \centering
    \includegraphics[width=\textwidth]{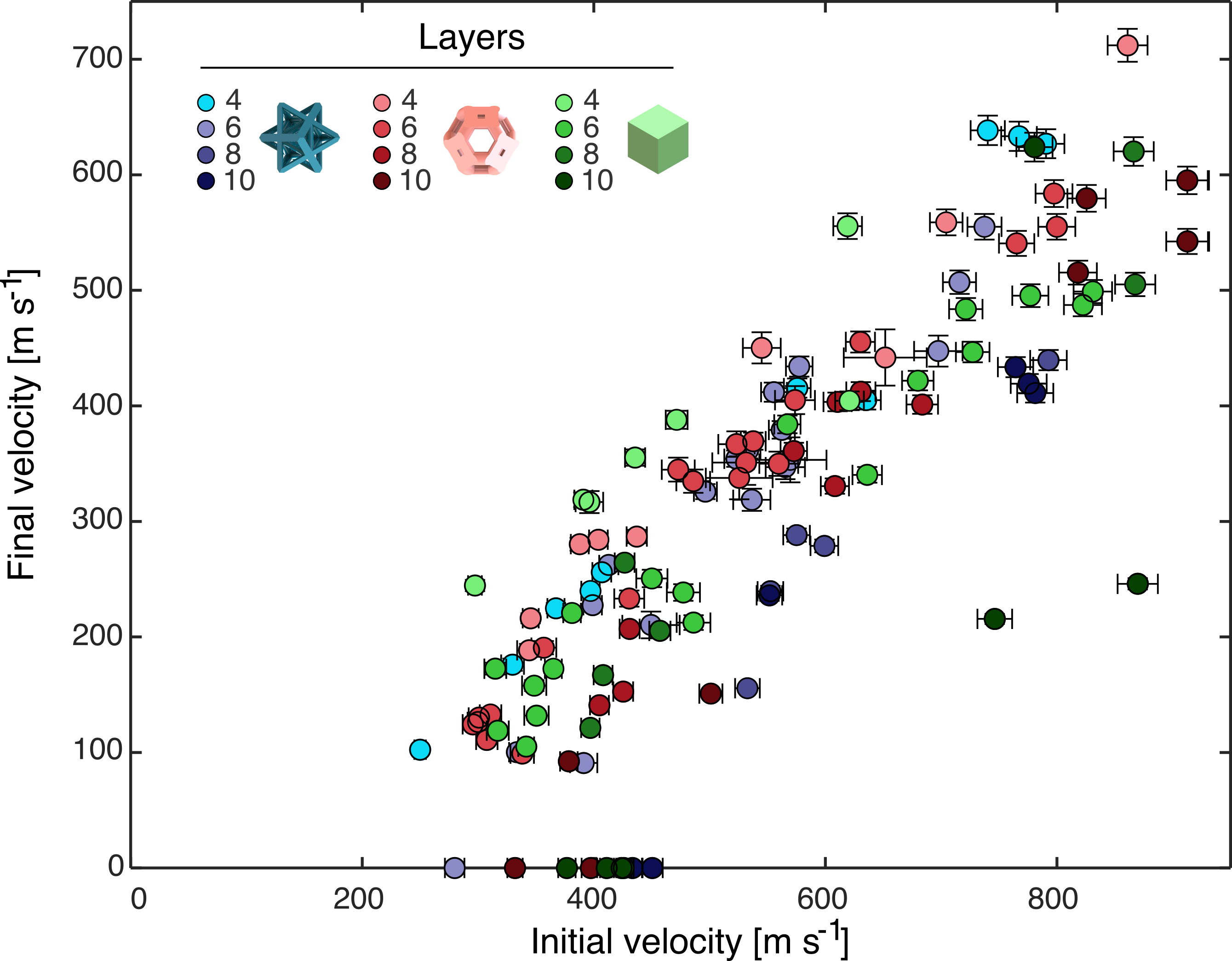}
    \caption{Initial and final velocities of $\sim$30 \textmu{}m diameter silica microspheres on single-architecture architected shields from 4 to 10 layer thickness. Layer nomenclature includes an implicit base layer identical across architectures.}
    \label{fig:ML_velo}
\end{figure*}

\begin{figure*}[t]
    \centering
    \includegraphics[width=\textwidth]{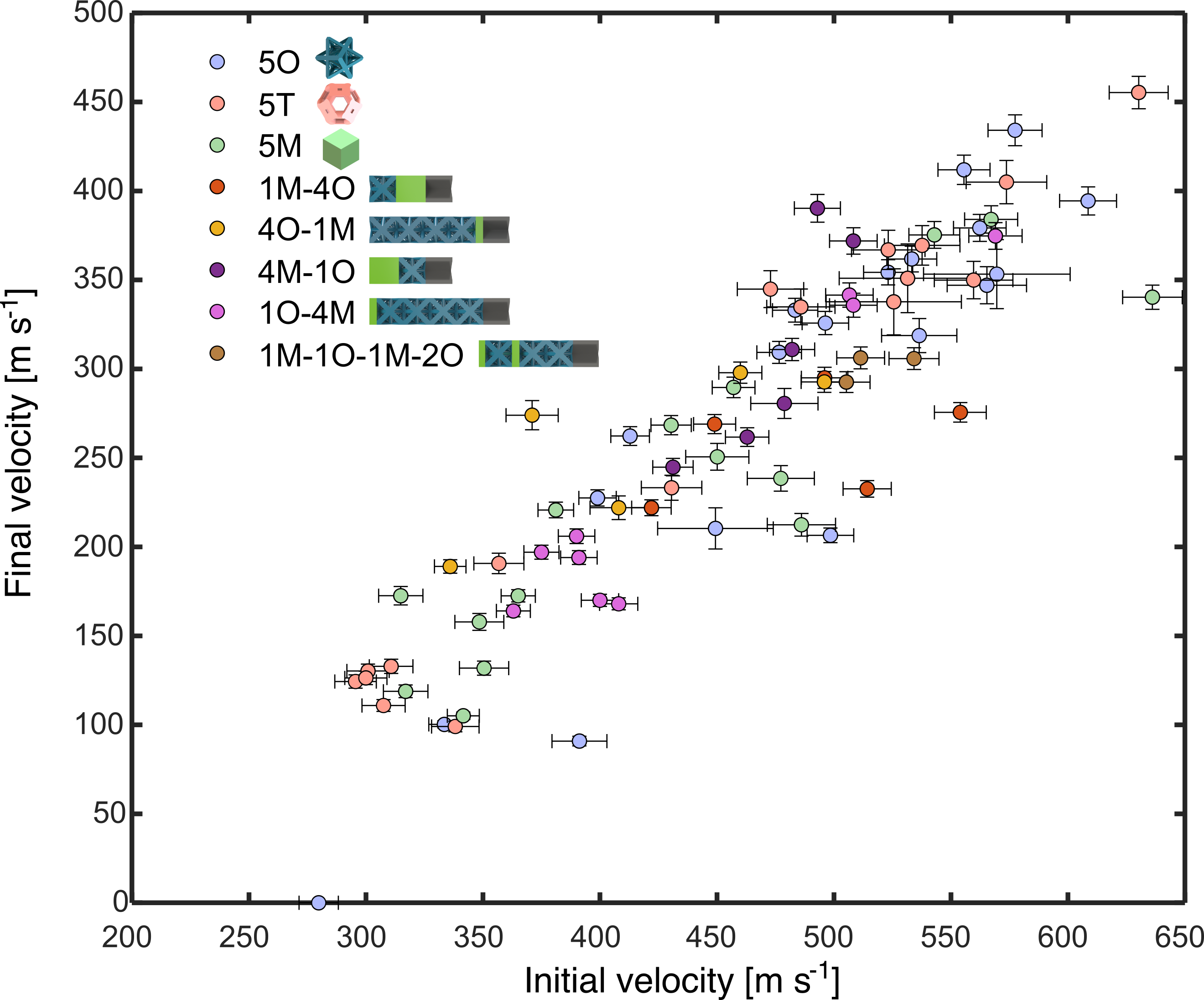}
    \caption{Incident and final velocities of $\sim$30 \textmu{}m diameter silica microspheres on six-layer architected shields. Architecture nomenclature includes an implicit, sixth base layer identical across architectures. Octet, tetrakaidecahedron, and monolithic suspended shield data was taken partially from Butruille et al. (2024) with permission.}
    \label{fig:HS_velo}
\end{figure*}

\begin{figure*}[t]
    \centering
    \includegraphics[width=\textwidth]{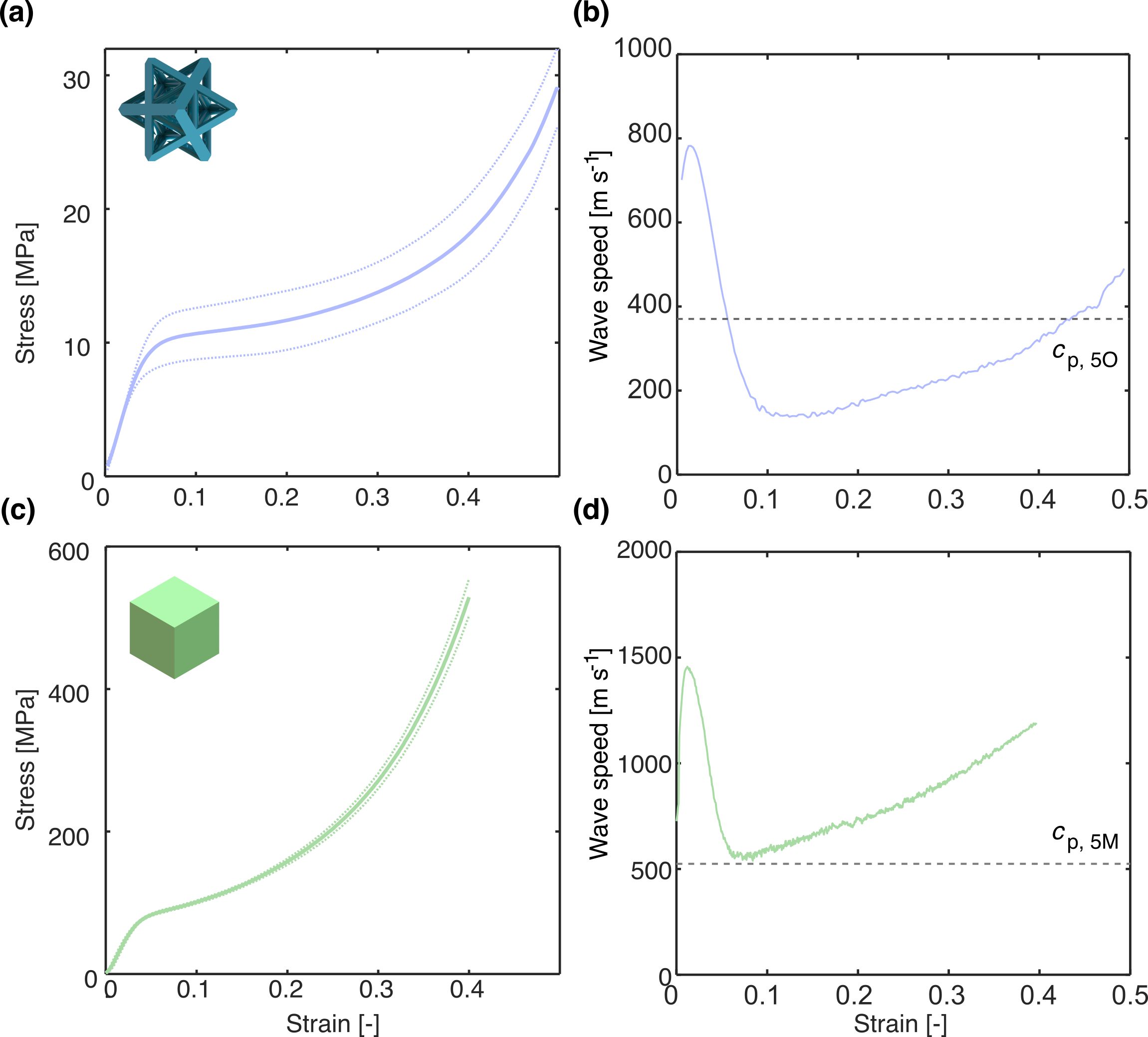}
    \caption{\textbf{a)} Uniaxial stress response for a 10$\times$10$\times$10 octet lattice with unit cells corresponding to those printed in the suspended shields. The solid line corresponds to the average response over three samples and the dotted line the standard deviation. \textbf{b)} Approximated octet wave speed as a function of compaction strain given by the expression $c_{p,~\textrm{octet}}=\sqrt{\frac{\partial\sigma/\partial\epsilon}{\rho_t}}$. The wave speed $c_{p,\textrm{5O}}$, found through fitting a propagating stress profile to the octet crater boundary in Figure 7 of the main text, is indicated by a dashed line. \textbf{c-d)} The uniaxial stress response of a monolithic IP-Dip pillar and the corresponding wave speed calculation, with $c_{p,\textrm{5M}}$ indicated by a dashed line. Uniaxial compression data was taken from Butruille et al. (2024) with permission.}
    \label{fig:SI_wavespeed}
\end{figure*}
\clearpage
\setcounter{table}{0}
\begin{table}
\centering
 \caption{Initial and final velocities across tomographically imaged shields}
  \begin{tabular}[htbp]{@{}lll@{}}

    \hline
    Architecture & $v_i$ [m s\textsuperscript{-1}] & $v_f$ [m s\textsuperscript{-1}] \\
    \hline
    5O  &  562 &  379 \\
    5M  &  636 & 340  \\
    1O-4M  &  390 & 206  \\
    4O-1M  &  449 & 269  \\
    4M-1O  &  463 &  262 \\
    1M-4O  &  496 & 295  \\
    1M-1O-1M-2O  &  505 &  293 \\
    \hline
  \end{tabular}
\end{table}

\end{document}